\documentclass[letterpaper,twocolumn,10pt]{article}
\usepackage{usenix-2020-09}

\usepackage[normalem]{ulem}
\usepackage{epsfig}
\usepackage{graphicx}
\usepackage{balance}
\usepackage{comment}
\usepackage{cite}
\usepackage{textcomp}
\usepackage{listings}
\usepackage{color,soul}
\usepackage{colortbl}
\usepackage{leading}
\usepackage{moreverb}
\usepackage{listings}
\usepackage{framed}
\usepackage{multirow}
\usepackage{pifont}
\usepackage{mathtools}
\usepackage{xstring,xifthen}
\usepackage{amsmath} 
\usepackage[utf8]{inputenc}
\usepackage{makecell}

\usepackage{tikz}

\makeatletter
\newif\if@restonecol
\makeatother

\usepackage[boxed]{algorithm2e}
\definecolor{lightgray}{gray}{0.9}
\definecolor{lightblue}{rgb}{0.9,0.9,1}
\definecolor{red}{rgb}{1,0,0}

  {\begin{list}{$\bullet$}%
     {\setlength{\parsep}{0pt}%
      \setlength{\topsep}{0pt}%
      \setlength{\itemsep}{2pt}}}%
  {\end{list}}

\newcommand\candidate[1]{} 

\newcommand{\cut}[1]{}

\newcommand\itemno[1]{({\em #1})}

\newcommand\mergerows[2]{\multirow{#1}{*}{\makecell[l]{#2}}}

\newcommand\tf[1]{\texttt{\textbf{\scriptsize#1}}}
\newcommand\tfi[1]{\texttt{\textbf{\small#1}}}

\newcommand\trust[3]{\StrLen{#2}[\twolen]\StrLen{#3}[\threelen]\ifthenelse{\twolen > \threelen}{\stackrel{\tf{#1} \textcolor{white}{\tf{s: #2}}}{\text{\Huge$\top$}^{\tf{s: #2}}_{\tf{w: #3}}}}{\stackrel{\tf{#1} \textcolor{white}{\tf{w: #3}}}{\text{\Huge$\top$}^{\tf{s: #2}}_{\tf{w: #3}}}}}

\newcommand\concattrust{\text{\huge$\cup$}}

\usepackage{subfigure}
\usepackage{titling}

\newcommand\sname{OctopOS\xspace}
\newcommand\Sname{OctopOS\xspace}

\newcommand\os{OS\xspace}

\newcommand\oses{OSes\xspace}

\newcommand\iodevice{I/O device\xspace}

\newcommand\iodevices{I/O devices\xspace}

\newcommand\ioservice{I/O service\xspace}
\newcommand\ioservices{I/O services\xspace}

\newcommand\TEE{TEE\xspace}
\newcommand\TEEs{TEEs\xspace}

\newcommand\splittrust{split-trust\xspace}

\newcommand\SplitTrust{Split-Trust\xspace}

\newcommand{\old}[1] {{\textcolor{blue}{}}} 

\newcommand\contribution{Equal contribution}

\newcommand\sectionspacebefore{\vspace{-3pt}}
\newcommand\sectionspaceafter{\vspace{-6pt}}
\newcommand\subsectionspacebefore{\vspace{-3pt}}
\newcommand\subsectionspaceafter{\vspace{-6pt}}

\begin{document}

\title{
Minimizing Trust
with Exclusively-Used Physically-Isolated Hardware
}

\date{}

\author{
{\rm Zhihao Yao$^{\dagger}\thanks{\contribution}$ , Seyed Mohammadjavad Seyed Talebi$^{\dagger}$$^{\textcolor{green}{\ast}}$, Mingyi Chen$^{\dagger}$}\\
{\rm Ardalan Amiri Sani$^{\dagger}$, Thomas Anderson$^{\ddagger}$}\\
$^{\dagger}$UC Irvine, $^{\ddagger}$University of Washington\\
{\normalsize\{z.yao, mjavad, mingyi.chen, ardalan\}@uci.edu, tom@cs.washington.edu}\\
}

\maketitle

\thispagestyle{empty}

\begin{abstract}

Smartphone owners often need to run security-critical programs on the same device as other untrusted and potentially 
malicious programs.
This requires users to trust hardware and system software to correctly sandbox malicious programs,
trust that is often misplaced.

Our goal is to minimize the number and complexity of hardware and software components that a smartphone owner needs to trust
to withstand adversarial inputs.
We present a multi-domain hardware design composed of statically-partitioned, physically-isolated trust domains.
We introduce a few simple, formally-verified hardware components to enable a program to gain provably exclusive and simultaneous
access to both computation and I/O on a temporary basis.
To manage this hardware, we present \sname, an \os composed of mutually distrustful subsystems.

We present a prototype of this machine (hardware and \os) on a CPU-FPGA board and show that it incurs a small hardware cost compared to modern SoCs.
For security-critical programs, we show that this machine significantly reduces the required trust compared to mainstream TEEs while achieving decent performance.
For normal programs, performance is similar to a legacy machine.

\end{abstract}

\sectionspacebefore
\section{Introduction}
\sectionspaceafter

Because of their ubiquity and portability, modern smartphones are often used to
run security-critical programs along with diverse, untrusted,
and potentially malicious programs. For example, most of us 
perform financial tasks, such as banking and 
payments~\cite{mobile_banking} on our smartphones.
Many of us also run health-related programs, e.g., to receive test results 
and diagnoses from our health providers.
There is also interest in using these devices to perform life-critical
tasks such as controlling an insulin pump~\cite{omnipod} or monitoring 
breathing~\cite{apnea}, although security concerns currently pose a roadblock~\cite{omnipod}.

Realizing this computing paradigm should be
straightforward.
The job of an operating system (OS) is
to isolate security-critical programs from other programs running
on the same hardware.
Yet, this has proven to be challenging in practice due to vulnerabilities in system software (e.g., \os, hypervisor, and device drivers)~\cite{VanderStoep2016, Linux_kernel_CVEs, syzkaller_bugs_found, Windows_10_CVEs, Xen_CVEs, Zhang2015, Palix2011, Ball2006, Chou2001} and hardware (e.g., processor, memory, interconnects, and I/O devices including their firmware)~\cite{Kim2014, Lipp2018, Kocher2019, van2018, Paccagnella2021, Weber2021, CVE-2021-0200}.
Malicious programs
can exploit these vulnerabilities to take control of the machine and any program running on it.
We must trust that hardware and system software
can effectively sandbox and neutralize malicious programs,
but this trust often proves to be misplaced.

To address this challenge, a new approach has emerged.
It uses \textit{Trusted Execution Environments (TEEs)} to host security-critical programs without requiring trust in the \os.
Unfortunately, 
today's \TEEs
still require us to trust the hardware
and the security monitor implementing the \TEE guarantees.
This trust has also proven unjustified. Existing TEEs have fallen victim to various attacks,
e.g., hardware-based side-channel attacks~\cite{Brasser2017, van2018, chen2019, Moghimi2020, Moghimi2017, Gotzfried2017, Schwarz2017,lipp2016, Zhang2016}, attacks exploiting software vulnerabilities~\cite{Cerdeira2020, OPTEE_CVEs, Breaking_Samsung_TrustZone, CVE-2015-6639}, and attacks based on design flaws~\cite{Hetzelt2017, Li2019_2, Wilke2020}.

In this paper, we present a solution to enable 
smartphones to be used for both security-critical and non-critical
programs.
Our goal is to minimize both the number and the complexity of hardware and software components that need to be \textit{strongly trusted} by the smartphone owner in order to execute a security-critical program.
As we will define in \S\ref{sec:trust_definitions}, we say that a component is strongly trusted if it needs to be able to withstand and neutralize adversarial inputs.

Our key principle is \textit{provably exclusive access} to hardware and software components.
That is, we design a solution to enable a security-critical program to \textit{exclusively use complex hardware and software components and be able to verify the exclusive use}.
Due to exclusive use, a component only needs to be \textit{weakly trusted}.
That is, it only needs to operate correctly in the absence of adversarial inputs.

More concretely, we present a hardware design for a smartphone.
Called a \textit{\splittrust hardware}, it comprises multiple trust domains, one or multiple for \TEEs, one for each \iodevice, one for a resource manager, and one for hosting a commodity \os and its programs.
The trust domains are \textit{statically-partitioned} and \textit{physically-isolated}: they each have their own processor and memory (and one I/O device in the case of an I/O domain) and do not share any underlying hardware components; they can only communicate by message passing over a hardware mailbox.
Moreover, we introduce a few simple, \textit{formally-verified} hardware components that enable a program to gain \textit{provably exclusive access} to one or multiple domains.

We then present \sname, an \os to manage this hardware.
Unlike existing \oses, which have a single, trusted-by-all nucleus, i.e., the kernel, \sname comprises mutually distrustful subsystems: a \TEE runtime for security-critical programs, \ioservices, a resource manager, and a compatibility layer for a commodity \os.

We rigorously evaluate the required trust, i.e., the Trusted Computing Base (TCB), of this machine.
We show that our machine significantly reduces the TCB compared to mainstream \TEEs and achieves one close to the lower bound.

We present a complete prototype of our machine (hardware and \os) on top of a 
a CPU-FPGA board (Xilinx Zynq UltraScale+ MPSoC ZCU102).
We use the powerful ARM Cortex A53 CPU to host the commodity \os (PetaLinux) and its programs with high performance.
We use the FPGA to build the other trust domains:
two \TEEs, a resource manager, and four I/O domains (an input domain, an output domain, a storage domain, and a network domain).
We use (weak) microcontrollers for these other domains, including the \TEEs.
This choice as well as the small number of \TEE domains is based on our observation that security-critical programs in smartphones, unlike regular programs, are often not as computationally intensive, and the number
of such programs that run simultaneously is typically small. 
In other respects, however, they are like normal programs: they 
start and stop, run in the background, do I/O, and so forth.

Using our prototype, 
we build two important security-critical programs for our machine:\footnote{We will open source our hardware design, \sname, security-critical programs, and formal verification proofs.}
\itemno{i} a banking program that can securely interact with the user, and
\itemno{ii} an insulin pump program that can securely execute its algorithm and communicate with (emulated) glucose monitor and pump.

We also use our prototype to measure the hardware cost and performance of our machine.
We show that the added hardware cost is small (i.e., 1-2\%) compared to modern SoCs.
Moreover, we show that \textit{security-critical program can achieve decent performance despite the use of weak microcontrollers for all TEE and I/O domains}.
Finally, we show that \textit{normal programs can achieve the same compute and I/O performance as on a legacy machine}.

\noindent\textbf{Secure hardware trend.}~~
Our vision of using physical isolation and exclusive use for security is in line with recent hardware trends from the smartphone industry.
Apple has integrated the Secure Enclave Processor (SEP) into its products~\cite{SEP} 
and used it to secure user's secret data
and to control biometric sensors (i.e., Touch ID and Face ID)~\cite{Face_Touch_ID}.
Similarly, 
Pixel 6 uses the tensor security core
to host security-critical tasks such as key management and secure boot~\cite{Pixel6_security}.
Our work takes this vision further by allowing programs (including those that rely on
I/O devices) to use dedicated processors by developing a model for how that can be safely done.

\sectionspacebefore
\section{Background}
\sectionspaceafter

\subsection{Trust Definitions}
\subsectionspaceafter
\label{sec:trust_definitions}

The hardware and software components that need to be trusted for a program to execute securely form its TCB.
We define two types of trust: \textit{strong trust} and \textit{weak trust}.
We say a component is strongly trusted if it needs to guard against \textit{adversarial inputs}.
For example, imagine an \os that is trusted to isolate a program from other malicious programs.
Malicious programs can issue adversarial syscalls to the \os concurrently to the protected program.
In such a case,
the component (e.g., the \os) must be trusted to prevent these other programs from exploiting any vulnerabilities (logical or implementation-related).
This is very challenging as demonstrated by the plethora of reported exploits.
Therefore, \textit{we believe that strong trust 
should be minimized for security-critical programs.}
We note, however, that there are methods for hardening hardware and software components, such as formal verification.
Strong trust is acceptable if a component is known to be adequately hardened against vulnerabilities.

We say that a component is weakly trusted if it just needs to operate correctly in the absence of adversarial inputs.
For example, consider the same \os mentioned above, but assume that the security-critical program is the only one running on top of the \os (and assume application-level networking).
In such a case, the component must only be trusted to
(1) not exert buggy behavior under normal usage, i.e., when processing well-formed inputs,
and (2) not be compromised by an adversary before use and upon distribution (e.g., through implanted backdoors).
These trust assumptions can be (more) easily met in practice by ensuring that: (1) component designers test it adequately under various expected usage models, (2) the source code of the component is available for inspection by security experts and users, and (3) users can verify the component before use through remote attestation.
Therefore, \textit{we believe that weak trust is acceptable for security-critical programs.} 
We also note that weak trust, as defined here, is the lower bound for trust as each component used by a security-critical program must at least be weakly trusted.

\subsectionspacebefore
\subsection{Trust in Existing Systems}
\subsectionspaceafter

\begin{figure}
\centering
\includegraphics[width=0.5\textwidth]{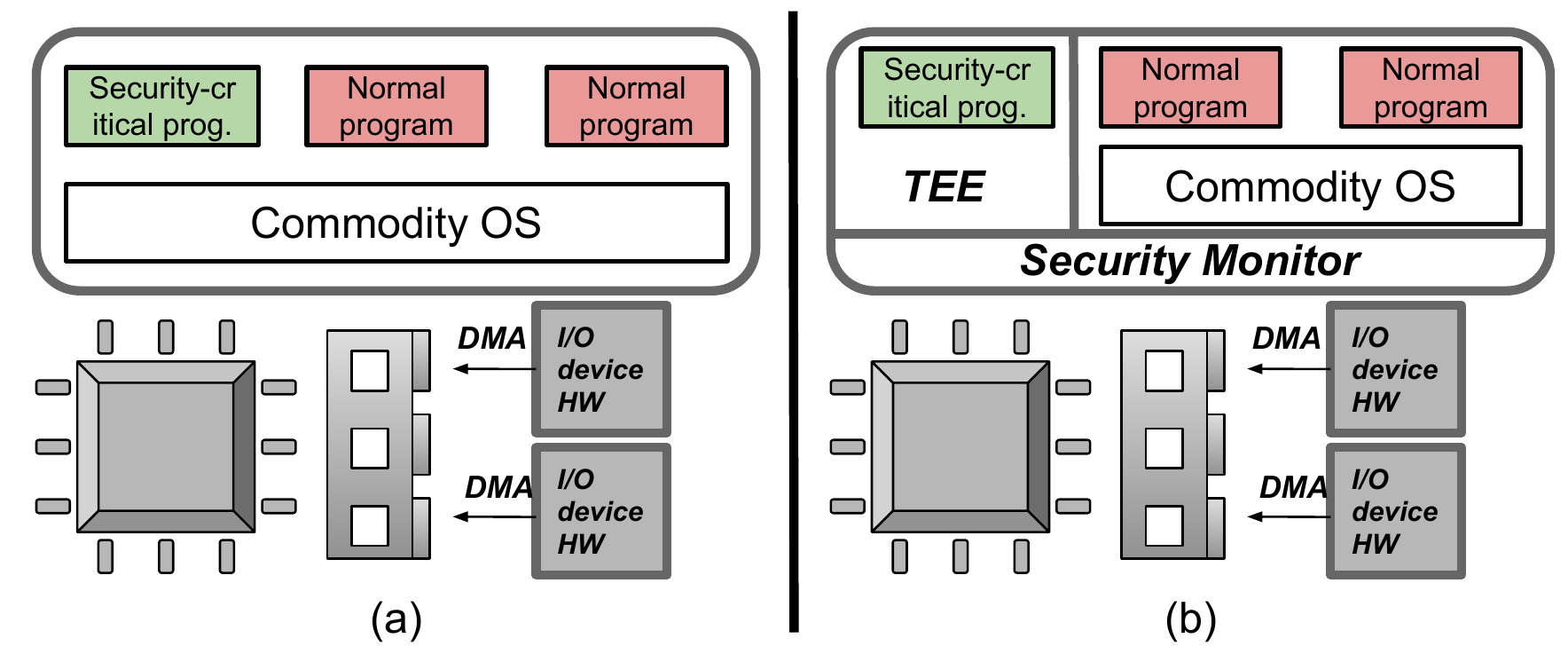}
\caption{\em (a) Traditional design where the \os isolates security-critical programs from normal programs. (b) Use of a \TEE to isolate a security-critical program.}
\label{fig:os}
\end{figure}

Historically, the \os has been a strongly-trusted part of the system (Figure~\ref{fig:os} (a)).
As commodity \oses have become more complex over the years, more and more vulnerabilities have been found in them, allowing malware to exploit them and compromise the \os~\cite{Linux_kernel_CVEs, VanderStoep2016, Windows_10_CVEs, syzkaller_bugs_found, Zhang2015, Chen2011, Palix2011, Ball2006, Chou2001}.
As an example, there have been about 1500 security vulnerabilities reported in the Linux kernel just since 2016~\cite{Linux_kernel_CVEs}.
Strong trust in commodity \oses is not warranted.

There have been several attempts to build trustworthy \oses.
These include
microkernels~\cite{Accetta1986, Liedtke1993, Gefflaut2000, Klein2009, Elphinstone2013}, 
exokernels and library \oses~\cite{Engler1995, Kaashoek1997, Porter2011, Baumann2015},
formally verified \oses (and hypervisors)~\cite{Klein2009, Gu2015, Gu2016, Vasudevan2016, Nelson2017, Sigurbjarnarson2018, Li2021, Li2021_2, Tao2021}, and \oses written in safe languages~\cite{Fahndrich2006, Hunt2007, Levy2017, Narayanan2020}.
While effective, these solutions require replacing commodity \oses with a new \os.
This is a challenging task due to the abundance of existing programs, device drivers, and developers for commodity \oses.
More importantly, using these \oses still requires strong trust in hardware, which is not warranted either, as we will discuss.

About two decades ago, a new approach started to gain popularity.
The idea is to create an isolated environment, called a \TEE, to host a security-critical program.
This allows the use of a commodity \os, but relegates it to be only in charge of untrusted, normal programs
such as games, utility apps, and entertainment platforms.
The \TEE enables a security-critical program to ensure its own integrity and confidentiality even if the \os is untrusted, but leaves the \os in charge of resource management (and hence the availability guarantee).
Figure~\ref{fig:os} (b) illustrates this design.
It shows a \textit{security monitor} is used to isolate a \TEE from the \os.
The security monitor can be implemented purely in software (i.e., a hypervisor)~\cite{Chen2008_Overshadow, Hofmann2013} or using a combination of hardware and software.
ARM TrustZone and Intel SGX are examples of the latter.
Others include
AMD Secure Encrypted Virtualization (SEV), Intel Trusted Domain Extensions (TDX), ARMv9's Realms ~\cite{ARM_CCA}, and Keystone for RISC-V~\cite{Lee2020}.

Despite their success, existing \TEE solutions 
still require many components to be strongly trusted including the security monitor and several hardware components such as the very complex processor, memory, I/O devices in some cases, and dynamically-programmable protection hardware such as address space controllers and MMUs.
Unfortunately, all of these components can be compromised by an adversary.
For examples, hypervisors contain many vulnerabilities~\cite{Xen_CVEs, Azab2016}.
\TEE \oses in TrustZone have also contained vulnerabilities and have been exploited in the past~\cite{Cerdeira2020, OPTEE_CVEs, Breaking_Samsung_TrustZone, CVE-2015-6639}.
AMD SEV has also been shown to contain several vulnerabilities due to design flaws
\cite{Hetzelt2017, Li2019_2, Wilke2020}.

Hardware components have been exploited as well.
Hardware-based side-channel attacks have recently emerged as a serious threat to computing systems.
For example, SGX enclaves and TrustZone have been compromised using several such attacks~\cite{Brasser2017, van2018, chen2019, Moghimi2020, Moghimi2017, Gotzfried2017, Schwarz2017, lipp2016, Zhang2016}.
The core reason behind this is that existing solutions execute the untrusted \os and \TEEs on the same hardware, forcing them to share underlying microarchitectural features such as cache~\cite{Brasser2017, lipp2016, Zhang2016, Moghimi2017, Gotzfried2017, Schwarz2017} and speculative execution engine~\cite{Lipp2018, van2018, Kocher2019, chen2019}, as well as architectural ones such as virtual memory~\cite{Moghimi2020}.
The memory subsystem has also proved vulnerable to Rowhammer attacks~\cite{Kim2014, Razavi2016, van2016, Xiao2016, Gruss2018, Loughlin2021}.
The complexity of these hardware components ensures that more vulnerabilities are likely to be discovered and exploited.
For example, researchers have recently demonstrated a suite of new side channels using the CPU interconnect~\cite{Paccagnella2021}, the x87 floating-point unit, and Advanced Vector extensions (AVX) instructions (among others)~\cite{Weber2021}.
\sectionspacebefore
\section{Key Goal and Principle}
\sectionspaceafter
\label{sec:key}

\noindent\textbf{Key goal.}~~
Our goal in this work is to minimize the number and complexity of strongly-trusted components.
It is difficult for complex hardware or software components to adequately protect themselves against adversarial inputs.
By contrast, simpler components can fend off adversarial inputs through comprehensive testing, analysis, and formal verification.

\noindent\textbf{Key principle.}~~
Our key principle to achieve this goal is \textit{provably exclusive access} to hardware and software components.
That is, we design our machine to enable a security-critical program to \textit{exclusively use complex hardware and software components and be able to verify the exclusive use}.
More specifically, our goal is to have most components, especially complex ones such as the processor and system software, (1) be reset to a clean state before use, (2) then used exclusively by a security-critical program in a verifiable fashion through remote and/or local attestation, and (3) then again reset to a clean state right after use.
In this case, such a component only needs to be weakly trusted as it does not need to worry about adversarial inputs while serving the security-critical program, nor does it need to worry about residual state from the security-critical program while serving other, potentially malicious, programs.

To realize this principle, we introduce a novel \textit{\splittrust hardware design} (\S\ref{sec:hardware}).
We then introduce an \os for this hardware, called \sname (\S\ref{sec:os}).

\sectionspacebefore
\section{\SplitTrust Hardware}
\sectionspaceafter
\label{sec:hardware}

Modern machines leverage hardware with a \textit{hierarchical privilege model}.
That is, hardware provides multiple privilege levels, each with more privilege than previous ones, with one all-powerful level to ``rule them all.''\footnote{A reference to Tolkien's The Lord's of the Rings.}
This model results inevitably in several complex, strongly-trusted components such as the processor, protection hardware, and system software.
 
\begin{figure}
\centering
\includegraphics[width=0.5\textwidth]{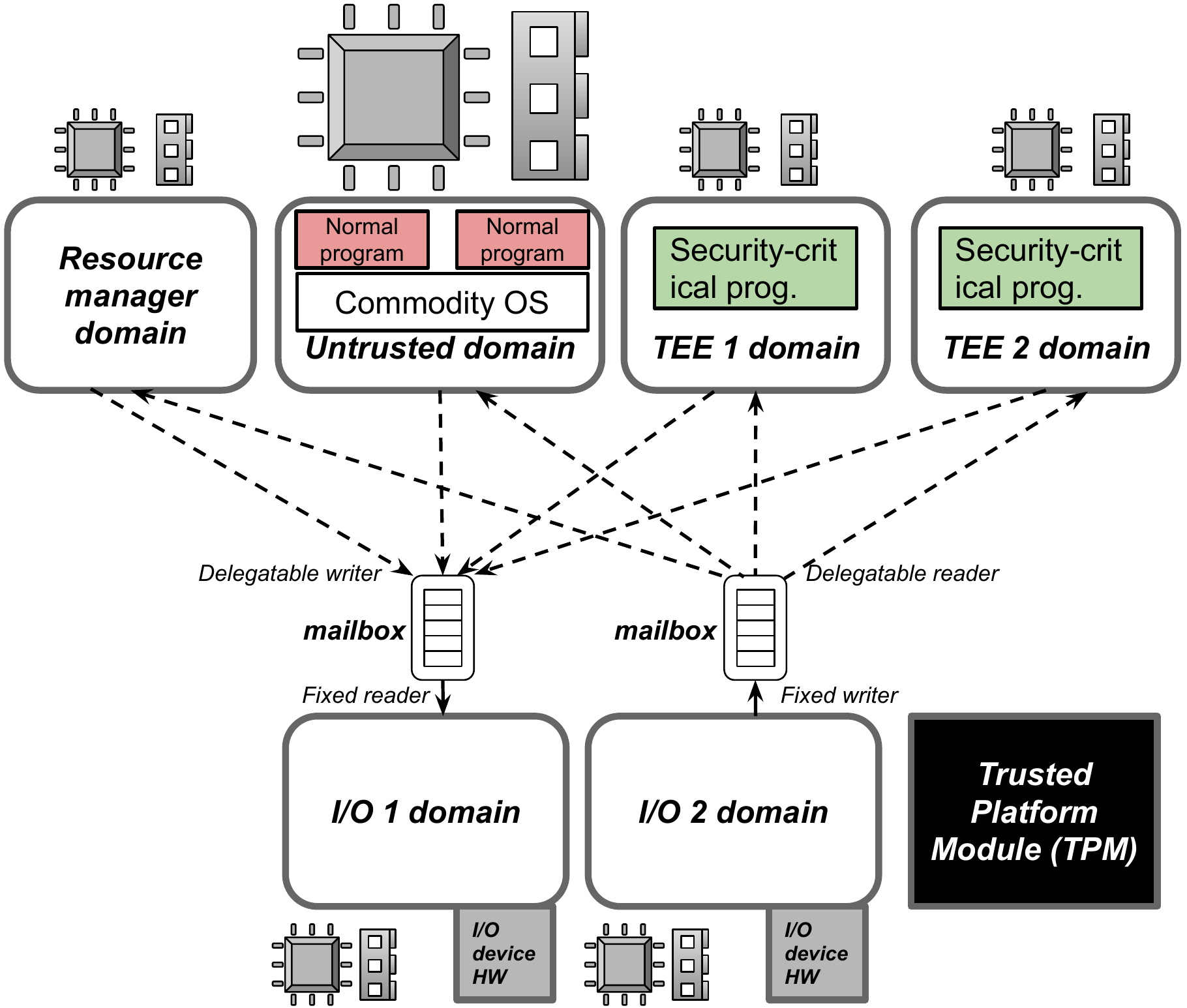}
\caption{\em Simplified overview of the \splittrust hardware. The figure does not show all mailboxes for clarity.}
\label{fig:overview}
\end{figure}

In this paper, we demonstrate a novel hardware design, the \textit{\splittrust} hardware, in which the hardware is split into multiple isolated trust domains.
Each domain is intended for one aspect of the machine:
one or multiple for \TEEs, one for each \iodevice (i.e., an I/O domain), one for a commodity \os and its untrusted programs (i.e., the
untrusted domain), and one for a resource manager, which is in charge of \textit{constrained} resource scheduling and access control.
The benefit of the \splittrust hardware is that a security-critical program can \textit{exclusively} take control of and use its own domain and \textit{exclusively} communicate with other domains (\S\ref{sec:interdomain}), e.g., for I/O and IPC, hence significantly reducing the strongly-trusted components.
Figure~\ref{fig:overview} shows a simplified view of this hardware design.
Next, we discuss its key aspects.

\subsectionspacebefore
\subsection{Physical Isolation and Static Partitioning}
\subsectionspaceafter
\label{sec:hardware_static}

We follow two important principles in our hardware design.
(1) Domains must be \textit{physically isolated} (i.e., share no hardware components).
(2) The isolation boundary between them cannot be programmatically and dynamically modified as \textit{there is no trusted-by-all hardware or software component}.
This implies that we cannot rely on programmable protection hardware, such as an MMU, IOMMU, or address space controller, to enforce isolation.
As a result, our design \textit{statically partitions} the hardware resources between domains.

More specifically, each trust domain has its own processor and memory.
We use a powerful CPU for the untrusted domain, which accommodates a commodity \os and its (untrusted) programs, to achieve high performance.
We use weaker microcontrollers for other domains in order to keep the hardware cost small.
Each domain has its own memory as well and domains do not (and cannot) share memory.

Importantly, each I/O domain also has exclusive control of an \iodevice, 
which is wired to and only programmable by the processor of that domain and which directly interrupts that processor. (We will discuss how DMA is handled in \S\ref{sec:hardware_dma}.)

\subsectionspacebefore
\subsection{Exclusive Inter-Domain Communication}
\subsectionspaceafter
\label{sec:interdomain}

To be able to act as one machine, the domains need to be able to communicate.
We introduce a simple, yet powerful, hardware primitive for this purpose: \textit{verifiably delegatable hardware mailbox}.
At its core, a mailbox is a hardware queue, allowing two domains (i.e., the writer and reader) to communicate through message passing.

The key novelty of our mailbox is how it enables exclusive communication using its \textit{delegation model}.
A mailbox has a fixed end (reader or writer) and a delegatable one.
The fixed end is hard-wired to a specific domain.
The delegatable one is wired to multiple domains, but only one can use it at a time, enforced by a hardware multiplexer within the mailbox.
This end is by default (i.e., after a mailbox reset) under the control of the resource manager domain.
But the resource manager can delegate it to another domain, which is then able to \textit{exclusively} communicate with the domain on the fixed end of the mailbox.

\begin{figure}
\centering
\includegraphics[width=0.5\textwidth]{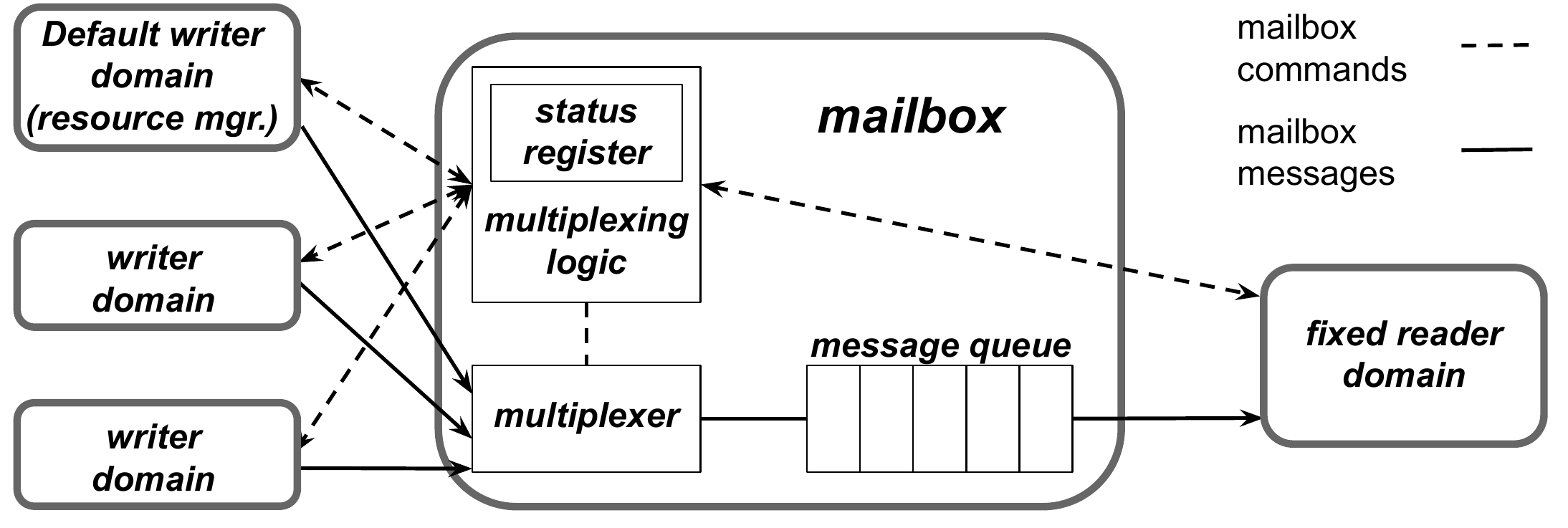}
\caption{\em Mailbox design.}
\label{fig:mailbox}
\end{figure}

Figure~\ref{fig:mailbox} shows the design of the mailbox with a fixed reader.
For example, consider the serial output domain in our prototype.
This domain is the fixed reader of a mailbox.
Any domain with write access to the mailbox can (exclusively) send content to the output domain to be displayed in the terminal.

The delegation model of our mailbox has another important property:
\textit{limited yet irrevocable} delegation.
When the resource manager delegates the mailbox to a domain, it sets a \textit{quota} for the delegation in terms of both the maximum number of allowed messages and maximum delegation time.
As long as the quota has not expired (i.e., \textit{a session}), the domain can use the mailbox and the resource manager cannot revoke its access to the mailbox.
The session expires when either the message limit or the time limit expires.
(The message limit can be set to infinite, but not the time limit.)

This delegation model enables a limited form of availability, which we refer to as \textit{session availability}.
That is, a domain with exclusive communication access to another domain can be sure to retain its access for a known period of time or number of messages.
This is critical for some security guarantees on smartphones.
For example, a security-critical program can ensure that the User Interface (UI) will not be hijacked or covered with overlays when the program is interacting with the user~\cite{Chen2014, Yan2019}.
Or a security-critical program that has authenticated to and hence unlocked a sensitive actuator domain (e.g., insulin pump)
can ensure that no other program can hijack the session and manipulate the actuator.

As the resource manager is not trusted by other domains, the delegation must be \textit{verifiable}.
The mailbox hardware provides a facility for this verification.
As Figure~\ref{fig:mailbox} shows, all domains connected to the mailbox
can read a status register from the mailbox hardware.
The status register specifies the domain that can read/write to the mailbox and the remaining quota.
The domain with delegated access can therefore verify its access and quota.
(Other domains will receive a dummy value when reading the status register for confidentiality.)

\subsectionspacebefore
\subsection{Power Management}
\subsectionspaceafter
\label{sec:hardware_pmu}

Our mailbox primitive cannot, on its own, guarantee session availability.
This is because we need to ensure that during a session, the domains used by a security-critical program remain powered up (assuming there is adequate energy if battery-powered).

The Power Management Unit (PMU) normally takes commands from the resource manager.
The resource manager uses this capability to reset other domains when needed, e.g., reset a \TEE domain before running a new program, or apply Dynamic Voltage Frequency Scaling (DVFS) to manage the system's power consumption.
(We do not support DVFS for the domains in our prototype.
Hence, in the rest of the paper, we mainly focus on the reset interface, although similar principles can be applied to DVFS.)

However, the resource manager is not a trusted component; hence it may try to reset a domain during a session.
Therefore, we add a simple hardware component, called the \textit{reset guard}, for the reset signals,
which ensures that as long as the quota on a mailbox has not expired,
the domains on both sides of the mailbox cannot be reset,
hence ensuring session availability.

\subsectionspacebefore
\subsection{Hardware Root of Trust}
\subsectionspaceafter

A hardware root of trust is needed during remote attestation to convince the party in charge of a security-critical program of the authenticity of the hardware and the correctness of the loaded program.
We use a Trusted Platform Module (TPM) to realize the root of trust for the \splittrust hardware.

\noindent\textbf{Why TPM?}~~
TPM, as specified by the Trusted Computing Group (TCG), is a tamper-resistant security co-processor connected to the main processor over a bus~\cite{TPM2_library_specification}.
Traditionally, it provides security features for the machine as a whole, such as the measurements of the loaded software.
This makes TPM unsuitable for more fine-grained security features, such as remote attestation of a specific program.
As a result, in-processor \TEE solutions, such as SGX,
integrate the root of trust in the processor itself, tightly coupling it with various features of the processor (such as virtual memory and cache), further bloating the strongly-trusted processor.

Our key insight is that TPM can provide fine-grained security features for a \splittrust machine since different components of this \os run in separate domains.
This allows the machine to enjoy the security benefits of TPM without suffering from its main limitation.

To integrate TPM into a \splittrust machine, we need a different set of parameters 
from the ones found in existing TPM chips.
We omit the details due to space limitation.

\subsectionspacebefore
\subsection{High Performance I/O}
\subsectionspaceafter
\label{sec:hardware_dma}

By default, the data plane of I/O domains are implemented over mailboxes.
However, this raises a performance concern due to additional data copies (to and from mailbox).
While the performance overhead is acceptable for \TEE domains, it is not so for the untrusted domain.
An important hardware primitive that enables a legacy machine to achieve high I/O performance is DMA.
To safely use DMA in our machine, we introduce \textit{domain-bound DMA}, defined with the following two restrictions.
(1) The DMA engine is hard-wired to only read/write to the memory of the untrusted domain.
(2) The DMA engine can stream data in/out of the I/O device only when the I/O domain is used by the untrusted domain.

We achieve this with a simple hardware component called the \textit{arbiter}, which is a switch that decides if the data streams of the \iodevice is connected to a DMA engine or to a simple FIFO queue accessible to the I/O domain.

\subsectionspacebefore
\subsection{Domain and Mailbox Reset}
\subsectionspaceafter

Domains and their mailboxes need to be reset before and after use (\S\ref{sec:key}).
We reset the mailboxes directly in hardware upon delegation, yield, and session expiration.
We leave the resetting of the domains to the resource manager, albeit under the limitations enforced by the reset guard (\S\ref{sec:hardware_pmu}).
Even though the resource manager is untrusted, this does not pose a problem since the program can verify, using local and remote attestation through TPM as well as some measures provided by the domain runtime that (1) a domain has been reset, (2) it has not been used since last reset, (3) it will be reset after use and before use by other domains.

This verification is rather straightforward for a \TEE domain.
The bootloader, which is stored in a ROM and is part of the remotely attested root of trust (\S\ref{sec:tcb_lower_bound}), is tasked with fully cleaning all the state information in the domain upon reset.
Once the program is loaded in the domain, it takes exclusive control of resources.
The only way for the resource manager to take the domain back is to reset it (if allowed by the reset guard), which then triggers the bootloader to clean the state.
The verification process is less straightforward for I/O domains.
We will discuss that in \S\ref{sec:os_io}.

\subsectionspacebefore
\subsection{Usability Discussion}
\subsectionspaceafter

We argue that the exclusive use of hardware resources by security-critical programs in our machine does not cause usability problems for normal programs, for three reasons.
First, security-critical programs in smartphones already use some I/O devices exclusively. 
For example, the UI (display and touchscreen) is used exclusively (e.g., when using TrustZone-based Protected Confirmation~\cite{Protected_Confirmation}) due to its small form factor.

Second, the performance impact on other I/O types, such as networking and storage, can be minimal when security-critical programs use short sessions, e.g., a few seconds.
In \S\ref{sec:evaluation_performance}, we experimentally demonstrate this impact for storage.
Moreover, TCP network connection keepalives persist for tens
of seconds. Further, since smartphone network connections are 
frequently dropped during handoffs, most widely used applications 
transparently re-establish lost connections without user visible changes. 
Security-critical programs can be designed to initiate, 
use, and close their connections in a single session (a practice that 
we use in our own sample programs).

It is also possible to mitigate these issues using multiple I/O domains of the same type.
For example, all smartphones have both WiFi and cellular network interfaces.
One can imagine allowing normal programs to share and use one of these while security-critical programs use the other (through two separate I/O domains in our hardware).

Third, most security-critical programs rely on only a subset of 
the I/O domains.
For example, our insulin pump program mainly requires access to its sensors and a brief access to storage.
While this program is running, all other I/O domains, e.g., network, UI, and even storage, can be used by normal programs.

\sectionspacebefore
\section{\Sname}
\sectionspaceafter
\label{sec:os}

We introduce \sname, an \os to manage the \splittrust hardware.
Unlike existing \oses, which have an all-powerful trusted-by-all kernel, \sname is composed of \textit{mutually-distrustful components}.
These components include \ioservices for I/O domains, a runtime for \TEE domains, a resource manager, and a compatibility-layer for the untrusted domain.

\subsectionspacebefore
\subsection{I/O Services}
\subsectionspaceafter
\label{sec:os_io}

Each I/O domain runs a service to manage it.
The \ioservice incorporates the software stack needed to program and use an I/O device, e.g., device driver.
In addition, it provides an API that can be called (through messages) by any client domain, i.e., the domain that has exclusive access to the mailboxes of the corresponding I/O domain.

I/O services play a role in ensuring that they are reset before and after use by other domains, as discussed next.

\noindent\textbf{Non-restricted I/O devices.}~~
Let us first consider I/O devices that can be used by a security-critical program without any restrictions during a session, such as the serial output and network devices in our prototype.
For these devices, when a security-critical program asks for access to an \iodevice, the manager resets the corresponding I/O domain (which triggers the bootloader on the I/O domain to clean all the state information in it) and then delegates its mailboxes to the \TEE domain running the program.
The program first uses the mailboxes of the corresponding I/O domain to verify its exclusive access and the session quota.
It then uses the attestation report from TPM to verify the software loaded in the I/O domain.
It also verifies that the I/O domain has not been used since it has been reset.
The latter requires assistance from the \ioservice.
That is, upon receiving the very first message and before processing the message (in order to prevent exposure to adversarial inputs), the service further extends the domain PCR register in TPM with a constant.
This way, the report from TPM reveals whether the domain has been used or not.

Before the program's session expires or is yielded, the program also needs to ensure that the I/O domain will be reset again before use by any other domain.
We also achieve this with assistance from the \ioservice.
That is, before yield/expiration, the program calls an API in the \ioservice to disable message processing, after which the service becomes unusable until it is reset again.

\noindent\textbf{Restricted devices.}~~
Next, let us consider I/O devices that cannot be used freely by a security-critical program during a session and require the resource manager to enforce restrictions (i.e., fine-grained access control).
In our prototype, storage falls in this category because it contains data of other programs as well.
Even if the data are encrypted, they need to be protected if a general availability guarantee is needed (\S\ref{sec:threat_model}).
For these devices, we still ensure exclusive access to the domain during the session.
We also ensure reset after use.
However, we cannot ensure the domain is reset to a clean state before use.
This is because after reset, the resource manager needs to communicate with the I/O service to restrict its usage before delegating the domain's mailboxes to a \TEE domain.

We have carefully designed an API for such \ioservices.
The core of the API revolves around the notion of an \textit{I/O resource}.
For example, in the case of the storage service, each partition is a resource.
The API allows the manager to allocate resources and bind them to specific security-critical programs.
It also allows the program to authenticate itself in order to use the resource and to verify the status of the service.
We omit the details of the API due to space limitation.
Finally, we note that this design makes the storage service strongly-trusted (\S\ref{sec:threat_model}).

\subsectionspacebefore
\subsection{\TEE Runtime}
\subsectionspaceafter

In order to facilitate the development of security-critical programs, we have developed a runtime for \TEEs, which provides a high-level API.
A program can choose to use this runtime, or can use its own.

We provide several categories of functions in this API:
(1) Requesting and verifying access to other domains (\S\ref{sec:os_io});
this category also helps the program manage the remaining quota of mailboxes by calling a callback function upon quota updates.
(2) High-level abstractions for using \ioservices such as socket-based networking and terminal prints.
(3) Assistance with the TPM, e.g., to request a remote attestation report.
(4) Support for secure IPC between \TEE domains.
(5) Security-critical routines such as cryptographic primitives.

\begin{table*}[ht] \renewcommand{\arraystretch}{1.1}
\centering
\fontsize{8}{10}\selectfont
\begin{tabular}{| l | l |}
\hline
	Property & Proved theorems\\
\hline
\hline
\mergerows{4}{Exclusive\\access}
& Domains w/o exclusive access to mailbox cannot change which domain has exclusive access, nor the remaining quota.  \\ \cline{2-2}

& If a domain does not yield its exclusive access, its exclusive access is guaranteed as long as the quota has not expired.  \\ \cline{2-2}

& The domain with exclusive access to the mailbox can correctly read or write from/to the queue. 	 \\ \cline{2-2}

& The domains w/o exclusive access to the mailbox cannot read/write to the queue.  \\ \cline{2-2}
\hline                                                                                                                                   
\hline                                                                                                                                   
\mergerows{2}{Limited\\delegation}
& When given exclusive access, a domain cannot use the mailbox more than its delegated quota.   \\ \cline{2-2}

&  When the quota delegated to a domain expires, the domain loses exclusive access.   \\ \cline{2-2}

\hline                                                                                                                                   
\hline                                                                                                                                   
\mergerows{2}{Excl. acc-\\ess verif.}
& The domain with exclusive access can correctly verify its exclusive access and remaining quota.   \\ \cline{2-2}

& The domain on fixed end of mailbox can correctly verify domain with exclusive access on the other end and remaining quota.    \\ \cline{2-2}

\hline                                                                                                    
\hline                                                                                                    
\mergerows{3}{Default\\exclusive\\access}
& After reset, the resource manager domain has exclusive access by default.  \\ \cline{2-2}

& The resource manager domain does not lose its exclusive access unless it delegates it.  \\ \cline{2-2}

& When a domain loses exclusive access (yield/expiration), the exclusive access will be given to the resource manager domain.  \\ \cline{2-2}
\hline                                                                                                                  
\hline                                                                                                                  
\mergerows{2}{Confide-\\ntiality}
& Domains w/o excl. access cannot use mailbox's verif. interface to find out which domain has excl. access and remain. quota. \\ \cline{2-2}

& Upon delegation/yield/expiration, the data in the queue is wiped. \\ \cline{2-2}
\hline
\end{tabular}
\caption{\em Theorems we prove for our mailbox. Proving some of these theorems require proving multiple lemmas not listed here.}
\vspace{-0.1in}
\label{tab:mailbox_theorems}
\end{table*}

\subsectionspacebefore
\subsection{Resource Manager}
\subsectionspaceafter

At a high level, the resource manager is in charge of resource scheduling, access control, and system-wide, untrusted I/O functionalities.
More specifically, it performs the following three tasks.
First, it makes constrained scheduling decisions.
When a new security-critical program needs to execute, or when an existing one requests exclusive communication with another domain (for I/O or IPC), the manager checks the availability of resources,
grants the request, or blocks it until the resource is available.
Compared to schedulers in commodity \oses, scheduling in \sname is more restricted.
This is because the resource manager cannot preempt a domain 
as long as mailbox quotas have not expired (\S\ref{sec:interdomain}).

Second, the resource manager restricts the usage of some I/O domains to enforce fine-grained access control, as discussed in \S\ref{sec:os_io}.

Finally, the manager implements system-wide, untrusted I/O functionalities.
For example, as the manager is the initial client of the input and output domains, it implements the shell (i.e., the UI).
The UI, however, can be delegated to security-critical programs upon request.

\subsectionspacebefore
\subsection{Untrusted Domain's Compatibility Layer}
\subsectionspaceafter

In \sname, a commodity \os runs in the untrusted domain, and hence by definition manages its own processor and memory.
Yet, the commodity \os is not given direct control of \iodevices as they are managed by separate I/O domains.

We address this issue by developing a compatibility layer for the untrusted \os.
In our prototype, which uses PetaLinux, the compatibility layer consists of several kernel modules,
each pretending to be a device driver.
Transparent to Linux and its program, they communicate to the resource manager to get access to the I/O services' mailboxes (or to setup DMA) and then communicate to them.

\sectionspacebefore
\section{Prototype}
\sectionspaceafter
\label{sec:implementation}

We have built a prototype of the \splittrust hardware and \sname on the Xilinx Zynq UltraScale+ MPSoC ZCU102 FPGA board.
We use the Cortex A53 ARM processor on the SoC for the untrusted domain in order to achieve high performance for the commodity \os (PetaLinux) and its programs.
We use the FPGA to synthesize 7 simple Microblaze microcontrollers (i.e., no MMU and no cache): two \TEE domains, the resource manager domain, and four I/O domains (serial input, serial output, storage, and Gigabit Ethernet).
We leverage the (single-threaded) Standalone library~\cite{Standalone_doc} to program the microcontrollers.
We use the entirety of the main memory for the untrusted domain.
For other domains, we use a total of 3.2 MB of on-chip memory including some ROM for bootloaders and some RAM.
We run the TPM (emulator)~\cite{IBM_TPM} on a separate Raspberry Pi 4 board connected to the main board through serial ports.
We use another Microblaze microcontroller to mediate the communications of the domains with the TPM.

In addition, we use the FPGA to synthesize the mailboxes (12 in total), the arbiter for DMA for the network domain (other domains do not support DMA), the reset guard, as well as 11 hardware queues for permanent domain connections (such as for all domains to communicate with TPM or for \TEE domains to communicate with the resource manager).
We note that we synthesize two types of mailboxes: \textit{control-plane mailboxes} and \textit{data-plane mailboxes}.
The former has 4 messages of 64 B each and the latter has 4 messages of 512 B each.
As a concrete example, our storage domain has 4 mailboxes: two for its control plane (send/receive) and two for its data plane (send/receive).

As mentioned in \S\ref{sec:hardware_static}, an I/O device is only programmable by its domain.
This includes access to registers and receiving interrupts from the I/O device.
In our prototype, we use I/O interrupts only for the network device and use polling for the rest.
The interrupts to the network domain's microcontroller is from the FIFO queue that holds the packets and are only used when the domain serves a \TEE domain (\S\ref{sec:hardware_dma}).
When serving the untrusted domain, the domain-bound DMA engine directly interrupts the A53 processor on DMA completion.

We faced
one noteworthy limitation in our prototype: the on-board SD card reader and flash memory are directly programmable by the A53 processor and hence could not be used for the storage domain.
Our solution was to connect a MicroSD card reader directly to FPGA through Pmod~\cite{Pmod_microSD}.
This provides physical isolation for the storage domain, but significantly degrades its performance due to Pmod's limited throughput.
Therefore, for performance evaluation, we instead use DRAM as our storage (we partition out a chunk of DRAM and use it exclusively for the storage domain).
This allows us to stress the performance of the mailboxes of the storage domain and get an upper bound for our storage performance, which we cannot do with the Pmod prototype.

We note that requiring an FPGA board to experiment with our machine may pose a road block for many researchers.
Therefore, we also develop an emulator for our hardware design.
The emulator runs on a Linux-based host \os such as Ubuntu and
is able to fully boot and run \sname.

Overall, we have implemented \sname and our hardware emulator in about 39k lines of C code (including 5k of modified drivers from Xilinx and crypto libraries).
We report the LoC for our hardware below.

\subsectionspacebefore
\subsection{Verified Hardware Design}
\subsectionspaceafter
\label{sec:implementation_verification}

The \splittrust hardware has only four simple hardware components that are strongly trusted (\S\ref{sec:threat_model}):
mailbox, DMA arbiter, reset guard, and ROM (for bootloaders).
We have implemented these components in 1630 lines of Verilog code as well as 800 lines of Python code.

The simplicity of our strongly-trusted hardware components enables us to formally verify them.
We use SymbiYosys to perform formal verification~\cite{symbiyosys}.
SymbiYosys is a front-end for Yosys-based formal hardware verification flows.
We use the SMTBMC engine, which uses $k$-induction to formally verify safety features in hardware.
Table~\ref{tab:mailbox_theorems} shows the list of theorems we prove for our mailbox (we omit the rest due to space limitation).
Overall, we developed 3000 lines of SystemVerilog code for our hardware verification.
\sectionspacebefore
\section{TCB \& Security Analysis}
\sectionspaceafter

\subsection{TCB Notation}
\subsectionspaceafter

The owner of a smartphone, when running a security-critical program in it, trusts various hardware and software components: the TCB.
We introduce and use a simple, compact notation for TCB, discussed here with an abstract example:

\vspace{-0.2in}
\begin{align*}
\tf{Owner}\trust{G1,G2}{CompA(1), CompB(2)}{CompC(3)}\concattrust\trust{G3}{1,2,3}{}
\end{align*}
\vspace{-0.2in}

The key operator is the $\top$ sign, which resembles a T (as in Trust).
It helps denote a \textit{set of trust assumptions}.
The elements on top of the $\top$ sign, e.g., \tfi{G1}, are the security guarantees, e.g., confidentiality and integrity.
This allows for differentiating trust assumptions for different guarantees and combining them using the $\cup$ sign.
The elements in front of \tfi{s:} and \tfi{w:} name the strongly- and weakly-trusted components.
For succinctness,
we tag a repeating component with a number in parenthesis on its first appearance and use the number in other locations.

\subsectionspacebefore
\subsection{Lower Bound of TCB}
\subsectionspaceafter
\label{sec:tcb_lower_bound}

This can be achieved if the machine is dedicated to executing a security-critical program:

\vspace{-0.2in}
\begin{align*}
\tf{Owner}\trust{C,I,A}{Prog.,RoT}{Proc.,Mem.,I/O,P.HW,interconnects}
\end{align*}
\vspace{-0.2in}

\noindent where \tfi{C}, \tfi{I}, \tfi{A} stand for Confidentiality, Integrity, and Availability;
\tfi{P.HW} is the protection hardware (e.g., MMU and IOMMU);
and \tfi{RoT} stands for Root of Trust.

This shows that the owner at the very least needs to strongly trust the (security-critical) program and the RoT.
The strong trust in the program is fundamental: the program needs to protect itself against adversarial inputs, e.g., malicious network packets.
Note that the program in the TCB includes the runtime used by the program to interact with the hardware.

The strong trust in the RoT is also fundamental and stems from the fact that an adversary controlling the machine may try to fool the verifier of remote attestation by attempting to attack and compromise the RoT.
The strong trust in the RoT includes strong trust in the bootloader, the ROM used to store the bootloader, the hardware/firmware used for remote attestation, e.g., TPM, as well as the hardware vendor that certifies attestation reports.

\begin{figure*}
\begin{align}
\begin{aligned}
\tf{Owner}\trust{C,I}{Prog.(1),SM(2),Proc.(3),Mem.(4),Sec-I/O(5),interconn.(6),P.HW(7),RoT(8)}{}\concattrust\trust{A}{1,2,3,4,5,6,7,8,OS}{}
\end{aligned}
\label{trust:existing_tee}
\end{align}
\vspace{-0.2in}
\begin{align}
\begin{aligned}
\tf{Owner}\trust{C,I,As}{Prog.(1),mailbox(2),reset-guard(3),arbiter(4),RoT(5)}{Proc.(6),Mem.(7),I/O(8),interconnects(9)}\concattrust\trust{Ag}{1,2,3,4,5,RM,SD}{6,7,8,9}
\end{aligned}
\label{trust:ours}
\end{align}
\vspace{-0.2in}
\end{figure*}

\subsectionspacebefore
\subsection{TCB of Existing Systems}
\subsectionspaceafter

First, we consider a traditional system that uses an \os to provide isolation:

\vspace{-0.2in}
\begin{align*}
\tf{Owner}\trust{C,I,A}{Prog.,OS,Proc.,Mem.,I/O,interconn.,P.HW,RoT}{}
\end{align*}
\vspace{-0.2in}

This shows that the owner strongly trusts the hardware including the processor, memory, I/O devices, protection hardware, and interconnects.
Moreover, the \os is also strongly trusted, including device drivers.
In this case, the program includes the libraries used by the program to interact with the \os and hardware.

Next, we write the TCB for a popular \TEE solution for smartphones, TrustZone, in Formula \ref{trust:existing_tee}.
\tfi{SM} is the security monitor (i.e., the secure world \os and monitor code).
We note that TrustZone allows the secure world to take full control of an I/O device, i.e., secure I/O (\tfi{Sec-I/O}).
Yet, this device and its driver are exposed to multiple programs in the secure world and hence are strongly trusted.
Another noteworthy issue is that the \os is strongly trusted when availability is needed as it is in charge of resource scheduling.

\subsectionspacebefore
\subsection{Our TCB}
\subsectionspaceafter
\label{sec:threat_model}

Formula \ref{trust:ours} shows the TCB of our machine.
\tfi{As}, \tfi{Ag}, \tfi{SD}, and \tfi{RM} stand for session availability, general availability, storage domain, and resource manager, respectively.
Our system requires strong trust in a few cases that were not part of the lower bound.
First, for confidentiality, integrity, and session availability, the owner needs to strongly trust the mailboxes used by the program, the arbiter (if domain-bound DMA is used), and the domain reset guard
as these components interact with untrusted components.
As discussed in \S\ref{sec:implementation_verification}, the simple design of these components allowed us to formally verify them, making this strong trust acceptable (\S\ref{sec:trust_definitions}).
Second, if a program needs general availability guarantees (e.g., it needs to be executed in fixed intervals)
and needs to store data across sessions, it needs to strongly trust the resource manager domain and the storage domain.
The only way to eliminate the strong trust in the storage domain for general availability is to have separate storage devices for each security-critical program.
Unfortunately, this is prohibitively expensive.

It is noteworthy that our machine eliminates the need to strongly trust several complex hardware and software components such as the processor, memory, I/O devices, the interconnects (since our machine does not share any busses between trust domains) and system software (security monitor, \os, and device drivers), compared to existing \TEEs.
Moreover, the hardware component listed as weakly-trusted (processor, memory, I/O, and interconnects) are those of the domains used by the security-critical program.
This has important implications: for example, it means that the complex powerful processor of the untrusted domain is not trusted at all (not even weakly).
Overall, the TCB of our machine is significantly smaller than modern, popular TEEs.
Moreover, our TCB is rather close to the lower bound.
Achieving a smaller TCB for a machine that can host security-critical and untrusted programs concurrently would be challenging.

\subsectionspacebefore
\subsection{Security Analysis}
\subsectionspaceafter

\noindent\textbf{Threat model.}
We assume an adversary can run untrusted and security-critical programs in the machine and tries to exploit any software or hardware vulnerabilities.
Below, we discuss various such attacks and their implications.
Physical attacks are out of scope.

\noindent\textbf{Software vulnerability-based exploits.}~~
Vulnerabilities in strongly-trusted software components would lead to attacks.
An attacker that compromises the program can obviously change its behavior.
An attacker that compromises the bootloader (including the code that cleans up the state in a domain upon reset) can falsify the remote attestation report or access/impact data from other sessions.
An attacker that can compromise the storage service can delete the program's data.
An attacker that can compromise the resource manager can starve the program of resources
(but cannot impact the availability of a given session once it is granted).
An attacker that manages to compromise other software components, e.g., \ioservices, other security-critical programs, and the untrusted \os, cannot mount an attack on the program.

\noindent\textbf{Hardware vulnerability-based exploits.}~~
In a \splittrust machine, unlike existing \TEEs, vulnerabilities in many complex hardware components such as the processor cannot be exploited since the adversary never shares the underlying hardware with the security-critical program.
Therefore, the attacker cannot leverage various hardware-based attacks such as cache side-channel attacks, interconnect side-channel attacks, speculative execution attacks, and Rowhammer attacks.
Only vulnerabilities in the strongly-trusted hardware components (i.e., mailbox, arbiter, reset guard, ROM, and TPM) would lead to attacks.
The first four of these are formally verified (\S\ref{sec:implementation_verification}) and TPM is a mature and secure technology.

\noindent\textbf{Timing side-channel attacks.}~~
All strongly-trusted software and hardware components are vulnerable to timing side-channel attacks.
In our machine, the only components that may expose useful timing channels are the TPM and the program runtime.
Such attacks (and others) have been demonstrated on TPMs before~\cite{Kauer2007, Sparks2007, Butterworth2013, Han2018, Moghimi2020_2}.
As TPM is a mature technology, vulnerabilities get fixed.
Indeed, there have been several works that formally verify various aspects of the TPM standard~\cite{Chen2013_2, Shao2015, Wesemeyer2020}.
We have not analyzed the timing channel of the runtime we have developed for security-critical programs.

\noindent\textbf{Power management attacks.}~~
These types of attacks can induce faults in the victim program's execution by manipulating the frequency or voltage of the processor and have been demonstrated against TEEs~\cite{Tang2017, Qiu2019, Murdock2020}.
As mentioned in \S\ref{sec:hardware_pmu}, our machine does not allow power management of a domain in a session, and hence mitigates such attacks.

Power management data can also be used as a side channel.
More specifically, an attacker man try to monitor the voltage and frequency of a domain (which changes according to DVFS) and use that as a side channel to extract secrets from a domain.
Indeed, the recent Hertzbleed attack uses this side channel to extract secret cryptographic keys~\cite{Wang2022}.

We note that our current prototype is not vulnerable to this side channel since our TEE domains do not support DVFS.
However, our hardware can support the use of DVFS-capable processors for TEE domains.
In such a case, we will need to close this channel.
To do so, we will need to ensure that the PMU does not leak any information about a domain to another domain.
This can be done rather trivially within the PMU firmware, which should be formally verified and hardened.

\noindent\textbf{Remote network attacks.}~~
Similar to a legacy machine, a security-critical program must protect itself against malicious network messages in our machine.
However, our machine provides some protection against network attacks that target the network stack.
This is because it sandboxes the network device and its device driver in its own domain.
As a result, programs that do not use the network at the time of a exploit are protected from these attacks.
This is in contrast to a legacy machine in which a single successful exploit of the kernel-based network stack may result in a full takeover.

\sectionspacebefore
\section{Sample Security-Critical Programs}
\sectionspaceafter
\label{sec:apps}

We briefly discuss two security-critical programs that we have built for our machine.

\noindent\textbf{I. Secure banking.}~~
Our secure banking program ensures that only the user can access confidential account information.
The program leverages several features of our machine.
First, it uses exclusive access to the UI (i.e., shell) to make sure all inputs come from the user (and not malware) and that outputs are only displayed to the user.
Upon getting exclusive access to the UI, the program needs to convince the user that they are interacting securely with the program.
It does so by displaying a secret established \textit{a priori} between the user and the bank.

Second, the program uses exclusive access to the network domain to transfer confidential information.
One might wonder why it is not adequate to use a secure networking protocol, such as TLS, for this purpose.
Such protocols leave open some side-channel attack vectors~\cite{Xiao2017}, which our exclusive network access closes 
against on-device attackers; external network side-channel attacks are still possible.
Finally, the program uses remote attestation to enable the bank cloud server to trust the program running on the user's device.

\noindent\textbf{II. Secure insulin pump.}~~
Diabetic patients need to monitor the glucose level in their blood and administer insulin accordingly.
New glucose monitor and insulin pumps have recently emerged that can perform this task automatically~\cite{omnipod}.

We build this security-critical program in our \os.
This program leverages exclusive access to the glucose monitor and insulin pump, e.g., through the headphone jack or via Bluetooth.
This way, the program can securely authenticate itself to the these devices and not worry that the session may be hijacked.
The program also uses exclusive access to the network domain to retrieve authentication tokens from the health provider's server and uses remote attestation to enable the provider's server to trust the program.
Finally, this program needs to be executed in fixed intervals
and store its sensor readings across sessions.
For this, it trusts the resource manager and the storage domain, as discussed in \S\ref{sec:threat_model}.
\sectionspacebefore
\section{Evaluation}
\sectionspaceafter

\subsection{Hardware Cost}
\subsectionspaceafter

We calculate an estimate for the number of transistors needed for our additional hardware components (all the components synthesized on the FPGA in our prototype).
We calculate this estimate by measuring the number of look-up tables, flip flops, and block RAMs used by our hardware and converting them to transistor count using the following estimates: 6 NAND gates per look-up table~\cite{gates_in_FPGA}, 6 transistors per NAND gate~\cite{basic_digital_circuits}, 24 transistors for each flip flop~\cite{Shizuku2014}, and 6 transistor for each bit of on-chip memory (assuming a conventional 6-transistor SRAM cell~\cite{Athe2009}).
Our calculation shows that our machine requires about 166.4 M additional transistors (162 M of which are used for on-chip memory).
Table~\ref{tab:hardware_cost} shows the breakdown.
This compares favorably with the number of transistors used in modern SoCs.
For example, Apple A15 Bionic and HiSilicon Kirin 9000 use 15 B transistors~\cite{A15_bionic, Kirin_9000}.
This means that, if our solution is added to an SoC or implemented as a chiplet~\cite{chiplet}, 
the additional hardware cost would likely be 1-2\%.

\begin{table}
\centering
\fontsize{8}{10}\selectfont
\begin{tabular}{| l | l | l |}
\hline
FPGA resource & Count & Equivalent transistor count\\
\hline
\hline
Look-up table & 69,999 & 2,519,964\\
\hline
Flip flop & 63,188 & 1,516,512\\
\hline
Block RAM & 27,061,649 (bits) & 162,369,894\\
\hline
\end{tabular}
\caption{\em Cost of additional hardware in our machine.}
\vspace{-0.1in}
\label{tab:hardware_cost}
\end{table}

\subsectionspacebefore
\subsection{Performance}
\subsectionspaceafter
\label{sec:evaluation_performance}

We measure various performance aspects of our machine.
Note that all domains except the untrusted one use an FPGA with a 100 MHz clock.
The Ethernet controller IP uses an external 50 MHz clock.
Therefore, our results represent a lower bound on our machine's performance; we expect superior performance on ASIC.
We repeat each experiment 5 times and report the average and standard deviation.

\noindent\textbf{Mailbox performance.}~~
We measure the throughput and latency of communication over our mailbox.
For throughput, we measure the time to send 10,000 messages of 512 B over a data-plane mailbox.
For latency, we measure the round trip time to send a 64 B message and receive an acknowledgment over a control-plane mailbox.
We perform these experiments in two configurations: one for communication between the hard-wired ARM Cortex A53 (the untrusted domain) and an FPGA-based Microblaze microcontroller, and one for communication between two FGPA-based Microblaze microcontrollers.
Table~\ref{tab:mailbox_performance} shows the results.
One might wonder why the A53-Microblaze configuration achieves lower performance.
We believe this is because this configuration requires the data to pass the FPGA boundary, hence passing through voltage level shifters and isolation blocks~\cite{UltraScale_TRM}.
Moreover, the FPGA is in a different clock domain than A53.

\begin{table}
\centering
\fontsize{8}{10}\selectfont
\begin{tabular}{| l | l | l |}
\hline
Configuration & Throughput (MB/s) & Latency ($\mu$s)\\
\hline
\hline
A53-Microblaze & 7.07$\pm$0 & 18.2$\pm$0\\
\hline
Microblaze-Microblaze & 9.64$\pm$0.01 & 15.26$\pm$0.05\\
\hline
\end{tabular}
\caption{\em Mailbox performance.}
\vspace{-0.1in}
\label{tab:mailbox_performance}
\end{table}

\noindent\textbf{Storage performance.}~~
We measure the performance of our storage domain, which uses the mailbox for its data plane (i.e., no DMA).
To do so, we perform 2000 reads/writes of 512 B each.
We evaluate three configurations: a best-case configuration where the storage domain directly performs reads/writes (hence giving us an upper bound on the DRAM-based storage performance),
and two configurations where the untrusted domain or a \TEE domain uses the storage service over the domain's mailboxes.
Table~\ref{tab:storage_performance} shows the results.
They show that our mailbox-based storage domain can achieve decent performance (as can also be seen from our boot-time measurements reported below).
It also shows that the additional copies caused by the mailbox add noticeable overhead compared to the best-case scenario.
To further improve this performance for the untrusted domain, one can use domain-bound DMA for the storage domain.

\noindent\textbf{Network performance.}~~
We measure the performance of our network domain, which uses domain-bound DMA for high performance for the untrusted domain (\S\ref{sec:hardware_dma}).
We evaluate three configurations, similar to those used for storage experiments.
For measuring the throughput for the baseline and the untrusted configurations, we use iPerf; for round-trip time (RTT) measurements, we use Ping.
For the TEE configuration, we develop custom programs for measurements.
For all experiments, we connect the board to a PC, which acts as a server.
Table~\ref{tab:network_performance} shows the results.
They show that our domain-bound DMA is capable of matching the performance of a legacy machine.
Moreover, the network performance for a TEE is decent.

We believe, based on some tests that we have conducted, that it is possible to further improve the TEE network performance by about 10 X.
This is because, currently in the network domain, we add an artificial delay between accessing the mailbox and the network IP, which limits performance.
We do so to prevent data corruption, which
according to our extensive investigation, is caused by a bug in the Ethernet AXI IP from Xilinx (potentially the bug discussed in~\cite{Ethernet_IP_bug}).
Since the IP is closed source, we are not able to fix the bug.

\begin{table}
\centering
\fontsize{8}{10}\selectfont
\begin{tabular}{| l | l | l |}
\hline
Configuration & {\scriptsize Read throughput(MB/s)} & {\scriptsize Write throughput(MB/s)}\\
\hline
\hline
Best-case & 8.13$\pm$0.00 & 6.10$\pm$0.00\\
\hline
Untrusted dom. & 4.17$\pm$0.09 & 4.06$\pm$0.00\\
\hline
\TEE domain & 4.39$\pm$0.00 & 3.93$\pm$0.00\\
\hline
\end{tabular}
\caption{\em Storage performance.}
\vspace{-0.1in}
\label{tab:storage_performance}
\end{table}

\begin{table}
\centering
\fontsize{8}{10}\selectfont
\begin{tabular}{| l | l | l |}
\hline
Configuration & Throughput (Mbit/s) & RTT (ms)\\
\hline
\hline
Baseline & 943$\pm$0 & 0.17$\pm$0.01\\
\hline
Untrusted domain & 943$\pm$0 & 0.17$\pm$0.02\\
\hline
\TEE domain & 0.567$\pm$0.001 & 23.92$\pm$0.02\\
\hline
\end{tabular}
\caption{\em Network performance.}
\vspace{-0.1in}
\label{tab:network_performance}
\end{table}

\noindent\textbf{Boot time and breakdown.}~~
We measure the time it takes our system to boot.
All the boot images are transferred from the storage domain to their corresponding domains over domain mailboxes.
Due to presence of multiple domains, booting \sname from a partition in the storage service is a carefully choreographed dance.
We boot \sname from the boot images stored in the boot partition of the storage service (\S\ref{sec:os_io}), 
requiring steps taken by the bootloaders in each domain and the resource manager, as follows.
\begin{itemize}
\item
The first domain to boot is the storage service.
The bootloader of this domain reads the storage service image from the drive and loads it onto its processor.
\item
The next domain to boot is the resource manager.
The bootloader of this domain communicates with the storage service using its API called over its mailbox.
The bootloader uses a simple file system (which is used to create the boot partition) to read the right blocks of data containing the resource manager image.
\item
Once the resource manager is booted, it assists all other domains to boot.
To do so, it invokes the storage service API in order to transfer the required images onto its data plane.
It then delegates the data plane to the corresponding domain, the bootloaders on which can receive the images and load them.
\end{itemize}

We note that the boot process includes extending the measurements of the boot images to the corresponding TPM PCRs as well.
This is mainly done by the bootloaders
, which are currently the root of trust in our prototype.
We have implemented a ROM for each domain to store its bootloader
, programmed as a part of the hardware bitstream.
Our measurements show that it takes 4.03$\pm$0.00 s to boot all domains excluding the untrusted domain, which takes an additional 8.65$\pm$0.01 s to boot.

\noindent\textbf{Untrusted program performance.}~~
We use the network file system to evaluate the performance of an untrusted program.
Our benchmark reads 100 files each containing 10,000 random numbers from a network file system, sorts them, and writes them back to the same file system.
We choose this benchmark since it stresses CPU, memory, and network (for which we have domain-bound DMA).
Our evaluation shows the benchmark takes the same amount of time (3.86$\pm$0.03 s) on our machine as it takes on a legacy machine with the same A53 processor, RAM, and Gigabit Ethernet (3.84$\pm$0.04 s).

\noindent\textbf{Security-critical program performance.}~~
We measure the execution time of a security-critical program.
The program reads a 1 MB file from the storage domain, computers its hash, and sends the hash over the network to a server.
Our measurements show that the overall execution time, when no other domain needs and hence competes for the storage and network domains, is 2.37$\pm$0.00 s.
Looking at the breakdown, this program takes
0.94$\pm$0.00 s 
to launch (including time needed to acquire exclusive access to storage and network, excluding local attestation through TPM), 
0.22$\pm$0.00 s to read the file from storage, 1.21$\pm$0.00 s to compute the hash, and less than 0.001 s to send the hash over the network.
(We however note that it is possible that the program might need to wait for the storage and/or network domains to become available if they are being used by other domains, e.g., the untrusted domain.)
To better asses this execution time, we write a normal program to perform similar tasks on a legacy machine with the A53 processor, RAM-FS, and Gigabit Ethernet.
This program takes 0.23$\pm$0.00 s to execute.

\noindent\textbf{Impact of exclusive I/O use.}~~
We evaluate the impact of executing a security-critical program that uses storage
on the storage
performance of the untrusted domain.
More specifically, we launch a security-critical program in a \TEE that exclusively reads 1 MB from and writes 1 MB to storage, while the untrusted domain is reading
a 100 MB file data (which normally takes 24.26$\pm$0.31 s
to finish).
Our measurements show that the security-critical program causes a 2.58$\pm$0.03 s gap where the untrusted domain cannot access the storage.
\section{Thoughts on Scalability and Performance of \TEEs}

The exclusive use of \TEE domains limits the number of concurrent security-critical programs.
Moreover, our choice of using weak microcontrollers, small amounts of memory, and I/O without DMA for \TEEs limits the performance of security-critical programs.
We believe that the former is not a serious issue since we do not expect a large number of security-critical programs executing simultaneously in a personal computer.

The latter is mostly a non-issue either since security-critical programs are more concerned with security guarantees than performance.
However, there are exceptions, for example, authentication of the user by applying machine learning algorithms to photos taken by the camera.
We believe that these programs can leverage accelerators (which will be available in the machine in the form of additional I/O domains).
Indeed, Nider et al. propose a machine with no CPU and several self-managed devices~\cite{Nider2021}, showing the diminished role of CPU for performance.
\sectionspacebefore
\section{Related Work}
\sectionspaceafter

\noindent\textbf{Physical isolation.}~~
Notary~\cite{Athalye2019} safeguards approval transactions by running its agent on a separate SoC from the ones running the kernel and the communication stack.
Our work shares the idea of using physically-isolated trust domains and also resets the domains before and after use by other programs.
In contrast, we show how to safely mediate access to shared I/O devices for a workload of concurrent security-critical and untrusted programs. 

Likewise, I/O-Devices-as-a-Service (IDaaS) suggests that I/O devices should have their own separate microcontrollers (and observes that they often do) and advocates 
for hardening their interfaces against potentially malicious kernel behavior~\cite{AmiriSani2019}.
Our approach also
uses separate I/O microcontrollers but does not require strong trust in the microcontroller software, by resetting
the I/O domain between uses.

\noindent\textbf{Exclusive use.}~~
Flicker~\cite{McCune2008} uses the late launch feature of Intel Trusted Execution Technology (TXT)~\cite{TXT_book},
to exclusively run a program on the processor.
The exclusive use of the hardware results in minimizing the strongly-trusted components.
However, Flicker's design requires stopping all other programs (including untrusted ones) when running a security-critical program.
Our approach can run untrusted programs and security-critical programs concurrently (albeit with the limitation that I/O domains cannot be shared). 
Consider our secure insulin pump program (\S\ref{sec:apps}),
which might need to be run frequently while the user is actively
doing other, less security-critical, tasks on the main processor.

\noindent\textbf{Secure I/O for TEEs.}~~
SGXIO uses a hypervisor and a TPM to create a trusted path for an SGX enclave to access an I/O device~\cite{Weiser2017}.
The solution requires the enclave program not only to trust SGX's firmware and hardware, but also the hypervisor.
\textsc{Cure}
\cite{Bahmani2021} adds a few hardware primitives in order to allow the security monitor to assign a peripheral (i.e., access to MMIO registers and DMA target addresses) to an enclave.
These primitives are designed to be programmed by a trusted-by-all security monitor (unlike our work).

\noindent\textbf{Time protection.}~~
Ge et al. add time protection to seL4, which closes many of the available side channels in commodity processors~\cite{Ge2019}.
As the paper mentions, 
some processors do not provide mechanisms needed to close channels.
Moreover, channels using busses could not be closed, and they have recently been shown to be effectively exploitable~\cite{Paccagnella2021}.
Our approach of using completely separate hardware for security-critical programs addresses these concerns for these programs.
We do, however, note that our approach (as it stands) does not scale to support all (normal) programs, which may have their 
own security needs.
Therefore, we believe that time protection remains an important abstraction to be explored for when the same processor is
asked to host multiple programs.

\noindent\textbf{Other TEE solutions.}~~
Komodo is a verified security monitor that can create enclaves for security-critical programs~\cite{Ferraiuolo2017}.
Use of formal verification warrants the strong trust in the security monitor, but not the ARM processor that hosts both security-critical and untrusted programs.

Sanctum uses hardware modifications to RISC-V alongside a software security monitor to create isolated enclaves~\cite{Costan2016_2}.
Compared to SGX, Sanctum enclaves are protected against both cache and page fault side-channel attacks.
While this is important, Sanctum does not address other potential hardware vulnerabilities such as side channels through interconnects.

\sectionspacebefore
\sectionspacebefore
\section{Conclusions}
\sectionspaceafter

Smartphone owners expect to use their devices for a mixture of security-critical and ordinary tasks,
yet this requires strong trust that the hardware and system software is able to isolate those tasks from each other,
trust that is often misplaced.
Our goal in this work is to minimize the TCB when executing security-critical programs.
We present a hardware design with multiple statically-partitioned, physically-isolated trust domains,
coordinated using a few simple, formally-verified hardware components, along with \sname, an \os to manage this hardware.
We describe a complete prototype implemented on an FPGA and show that it incurs a small hardware cost.
For security-critical programs, our machine significantly reduces the TCB 
compared to existing solutions, and achieves decent performance.
For normal programs, it achieves similar performance to a legacy machine.

\section*{Acknowledgments}

The work was supported by NSF Awards \#1617513, \#1718923, \#1846230, and \#1953932.
The authors thank Felix Xiaozhu Lin for his invaluable feedback on an earlier draft of this paper.
They also thank Xilinx for donating an FPGA board to this project.

\bibliographystyle{plain}
\bibliography{ardalan}

\end{document}